\documentclass{sig-alternate-10pt}
\usepackage{algpseudocode,multirow}
\usepackage{url,subfigure,graphicx}
\usepackage{algorithm}
\begin{document}
%

\title{It Doesn't Break Just on Twitter. Characterizing Facebook content During Real World Events}

\numberofauthors{1} 
\author{
\alignauthor
Prateek Dewan, Ponnurangam Kumaraguru\\
       \affaddr{Indraprastha Institute of Information Technology - Delhi (IIITD)}\\
       \email{\{prateekd,pk\}@iiitd.ac.in}
}

\date{10 August 2013}

\maketitle
\begin{abstract}
Multiple studies in the past have analyzed the role and dynamics of the Twitter social network during real world events. However, little work has explored the content of other social media services, or compared content across two networks during real world events. We believe that social media platforms like Facebook also play a vital role in disseminating information on the Internet ~during real world events. In this work, we study and characterize the content posted on the world's biggest social network, Facebook, and present a comparative analysis of Facebook and Twitter content posted during 16 real world events. Contrary to existing notion that Facebook is used mostly as a private network, our findings reveal that more than 30\% of public content that was present on Facebook during these events, was also present on Twitter. We then performed qualitative analysis on the content spread by the most active users during these events, and found that over 10\% of the most active users on both networks post spam content. We used stylometric features from Facebook posts and tweet text to classify this spam content, and were able to achieve an accuracy of over 99\% for Facebook, and over 98\% for Twitter. This work is aimed at providing researchers with an overview of Facebook content during real world events, and serve as basis for more in-depth exploration of its various aspects like information quality, and credibility during real world events.
\end{abstract}



\section{Introduction}

Over the past decade, online social media has stamped its authority as one of the largest information propagators on the Internet. Nearly 25\% of the world's population uses at least one social media service today.~\footnote{\url{http://www.emarketer.com/Article/Social-Networking-Reaches-Nearly-One-Four-Around-World/1009976}} People across the globe actively use social media platforms like Twitter and Facebook for spreading information, or learning about real world events these days. A recent study revealed that social media activity increases up to 200 times during major events like elections, sports, or natural calamities~\cite{szell2014}. This swollen activity has drawn great attention from the computer science research community. Content and activity on Twitter, in particular, has been widely studied by researchers during real-world events~\cite{becker2011beyond,hu2012breaking,kwak2010twitter,sakaki2010earthquake,weng2011event}. However, few studies have looked at social media platforms other than Twitter to study real-world events~\cite{chen2009event,hille2013like,osborne2012bieber}. Surprisingly, there has been hardly any work on studying content and activity on Facebook during real world events~\cite{westling2007expanding}, which is five times bigger than Twitter in terms of the number of monthly active users.~\footnote{\url{http://www.diffen.com/difference/Facebook_vs_Twitter}}

Facebook is currently, the largest online social network in the world, having more than 1.28 billion monthly active users.~\footnote{\url{http://newsroom.fb.com/company-info/}} However, unlike Twitter, Facebook's fine-grained privacy settings make majority of its content private, and publicly inaccessible. About 72\% Facebook users set their posts to \emph{private}.~\footnote{\url{http://mashable.com/2013/07/31/facebook-embeddable-posts/}} This private nature of Facebook has been a major challenge in collecting and analyzing its content in the computer science research community. But even with a small percentage of content being public, this content (approx. 1.33 billion posts per day~\cite{fbwhitepaper:2013}) is much more than the entire Twitter content (approx. 500 million tweets per day).~\footnote{\url{http://www.sec.gov/Archives/edgar/data/1418091/000119312513390321/d564001ds1.htm}} This mere volume of the publicly available Facebook content makes it a potentially rich source of information. In addition, recent introduction of features like hashtag support~\cite{Facebook:2013} and Graph search for posts~\cite{Room:2013}, have largely increased the level of visibility of public content on Facebook, either directly or indirectly. Users can now \emph{search} for topics and hashtags to look for content, in a fashion highly similar to Twitter; thus making the public Facebook content more visible and consumable by its users. This increasing public visibility, and an enormous user-base, potentially makes Facebook one of the largest and most widespread sources of information on the Internet, especially during real world events, when social media activity swells significantly.

With such dominance in terms of volume of content and user base, it is surprising that Facebook content hasn't yet been studied a lot in the research community. In this work, we explore whether there even exists content related to real world events on Facebook? If so, are Facebook and Twitter cognate networks in terms of content during real world events? To what extent does their content overlap in terms of numbers and quality? To answer these questions, we present a comparative analysis between Facebook and Twitter content during 16 real-world events. 
Our broad findings and contributions are as follows:

\begin{itemize}

\item For 13 out of the 16 events analyzed, we found that 30.56\% of textual content that appeared on Facebook, also appeared on Twitter during real world events.

\item Real world events appear fairly quickly on Facebook. We found that on average, an unprecedented event appeared on Facebook in just over 11 minutes after taking place in the real world. For Twitter, this statistic has been previously reported as over 2 hours for generic events~\cite{Osborne:2014}.

\item Characterization and analysis of spam content on Facebook and Twitter. Using only stylometric features extracted from textual content posted on these networks, we were able to achieve accuracies of over 99\% and 98\% on Facebook and Twitter respectively.

\end{itemize}

To the best of our knowledge, this is the first attempt at comparing Facebook and Twitter content, and characterizing Facebook content at such a large scale during real world events. 
This work uncovers the potential of Facebook in disseminating information during real-world events. Knowing that Facebook has a high overlap with Twitter, and large amounts of information about real world events, government agencies can use Facebook as a platform to reach out to masses during critical events. Event-related content on Facebook can also be studied further by researchers to study information credibility on this network, and identify bad quality sources of information to prevent rumor propagation on Facebook during such events.

The rest of the paper is arranged as follows. Section~\ref{sect:relatedwork} discusses the related work; the data collection methodology is discussed in Section~\ref{sect:methodology}. Section~\ref{sect:analysis} presents our comparative analysis of Twitter and Facebook content. Qualitative analysis of content is presented in Section~\ref{sect:qual}, followed by spam analysis in Section~\ref{sect:spam}. Section~\ref{sect:last} summarizes our findings, and discusses the limitations.

\section{Related work} \label{sect:relatedwork}

A recent survey of 5,173 adults suggested that 30\% of people get their news from Facebook, while only 8\% receive news from Twitter and 4\% from Google Plus~\cite{Jesse-Holcomb:2013}. These numbers suggest that researchers need to look beyond Twitter, and study other social networks to get a better understanding of the flow of information and content on online social media during major real world events. Although there has been some work related to events on Wikipedia~\cite{osborne2012bieber} and Flickr~\cite{chen2009event}, there hardly exists any work on studying Facebook content during real world events. Osborne et al. recently examined how Facebook, Google Plus and Twitter perform at reporting breaking news~\cite{Osborne:2014}. Authors of this work identified 29 major events from the public streams of these three social media platforms, and found that Twitter was the fastest among them. Authors also concluded that all media carry the same major events. However, there was no analysis on the type and similarity of content that was found across these social media services. Our dataset is partially similar to the one used by authors in this work, but our objectives differ. While Osborne et al. performed a comparative analysis on these networks for finding which one does better at breaking news, our emphasis is on characterizing Facebook content, and comparing it with that of Twitter during real world events. Further, while Osborne et al. collected a random stream of data from Twitter and Facebook over 3 weeks (which, by itself, is a research challenge), and extracted events from this data; we took an opposite approach. We collected only event-specific data for major events spanning over a much longer period of 9 months. This approach gave us a richer and bigger set of \emph{event-specific} data (tweets and Facebook posts).

\paragraph{Real world events on Twitter}

In recent years, studying Twitter has been the main focus for researchers in context of real world events. Kwak et al.~\cite{kwak2010twitter} showed a big overlap between real-world news headlines and Twitter's trending topics. Authors of this work crawled the entire Twitter network, and showed that the majority (over 85\%) of topics on Twitter were headline news or persistent news in nature. This work triggered researchers in the entire computer science community to explore Twitter as a news media platform during real-world events. Other research also highlights the use of Twitter as a news reporting platform~\cite{java2007we,naaman2010really}. Petrovic et al. recently conducted a comparative study on 77 days worth of tweets and Newswire articles, and discovered that Twitter reports the same events as newswire providers, in addition to a long tail of minor events ignored by mainstream media. Interestingly, their findings revealed that, neither stream, i.e. Twitter or newswire, leads the other when dealing with major news events~\cite{petrovic2013can}. 


Twitter has been used widely during emergency situations, such as wildfires~\cite{de2009omg}, hurricanes~\cite{hughes2009twitter}, floods~\cite{vieweg2010microblogging} and earthquakes~\cite{earle2010omg,kireyev2009applications,sakaki2010earthquake}. Journalists have hailed the immediacy of the service which allowed ``to report breaking news quickly - in many cases, more rapidly than most mainstream media outlets''~\cite{poulsen2007firsthand}. Sakaki et al.~\cite{sakaki2010earthquake} explored the potential of the real-time nature of Twitter and proposed an algorithm to detect occurrence of earthquakes by simply monitoring a stream of tweets in real-time. Here, the authors took advantage of the fact that users tweet about events like earthquakes as soon as they take place in the real world, and were able to detect 96\% of all the earthquakes larger than a certain intensity.


The overwhelmingly quick-response, and public nature of Twitter, along with its tremendous reach across the globe have made it the primary focus for researchers to study. 
In this work, we try to bring out the relevance of Facebook during real-world events by comparing its content with data from Twitter. Given that Twitter has already been well-studied in this context, we believe that this would be a sound comparison to establish Facebook's grit as a content sharing platform during real-world events.

\section{Methodology} \label{sect:methodology}

We now discuss the methodology we used for collecting event specific data from Facebook and Twitter. We selected a combination of medium, and high impact events in terms of social media activity. Eight out of the 16 events we selected, had more than 100,000 posts on both Facebook and Twitter; 6 of these 8 events saw over 1 million tweets. 

\subsection{Event specific data collection}

We used the MultiOSN framework~\cite{dewan2013multiosn} for collecting event specific data from multiple social media platforms. MultiOSN uses REST based, keyword search API for collecting public Facebook posts, and Twitter's search and streaming APIs for collecting public tweets. The selection of events to capture, and related keywords was manual. A similar technique has been previously used by researchers, to collect event-specific data~\cite{gupta20131,hughes2009twitter}. The same keywords were used to collect data for both Facebook and Twitter, in order to avoid bias. Table~\ref{stats} presents the detailed statistics of our dataset, in chronological order of occurrence of the events. The detailed description of these events can be found in Table~\ref{tab:descofevents}.

\begin{table*}[!ht]
\begin{center}
    \begin{tabular}{l|l|l|l|p{1.6cm}|l}
    \hline
    Event Name                      & Duration                 & FB posts & FB users & Tweets & Twitter Users \\ \hline
    Indian Premier League           & Apr 1 - Jun 7         & 662,459              & 381,925               & 2,627,197    & 485,632              \\
    Iran Earthquake                 & Apr 16 - Apr 17      & 2,779                & 2,537                  & 171,232      & 113,602               \\
    Boston Blasts                   & Apr 16 - Apr 20      & 1,480,467            & 1,213,990               & 2,840,570    & 1,714,253              \\
    Kashmir Earthquake              & May 1 - May 6            & 1,043                & 1,002                  & 86,541       & 35,786                \\
    Mothers Day Shooting            & May 13 - May 14          & 5                    & 5                     & 8,632        & 7,050                 \\
    Oklahoma Tornado                & May 20 - May 31          & 1,394,855            & 992,378                & 2,393,058    & 1,388,769              \\
    London Terror Attack            & May 23 - May 31          & 86,083               & 68,966                 & 361,460      & 185,145               \\
    North India Floods              & Jun 14 - Aug 8       & 211,388              & 150,212                & 147,416       & 63,527                \\
    Champions Trophy                & Jun 16 - Jun 26        & 349,306              & 241,419                & 1,016,605    & 567,024               \\
    Birth of Royal Baby             & Jul 15 - Aug 18      & 90,096               & 74,174                 & 1,830,696    & 1,083,955              \\
    Creation of new Telangana state & Jul 31 - Aug 16      & 253,926              & 158,255                & 135,580      & 37,364                \\
    Washington Navy Yard Shooting   & Sep 16 - Oct 1 & 4,562                & 3,091                  & 349,768      & 173,267               \\

    Cyclone Phailin & Oct 12 - Oct 19      & 60,016              & 43,096                & 110,047      & 46,955                \\
    Typhoon Haiyan & Nov 8 - Nov 13      & 486,325  &  355,705 &  1,257,171 & 610,809\\
    Metro North Train Derailment & Dec 1 - Dec 6      &  1,165  & 1,036  & 28,315  & 16,830\\
    Death of Nelson Mandela & Dec 5 - Dec 22      &  1,318,854 &   1,109,783 &  4,104,463 &  2,077,719  \\ \hline
    \end{tabular}
\end{center}
\vspace{-10pt}
\caption{Total amount of data collected for the 16 events tracked; ordered chronologically.
}
\label{stats}
\vspace{-10pt}
\end{table*}


\begin{table*}
\small
    \begin{tabular}{p{3cm}|p{14cm}}
\hline
    Event                         & Description                   \\ \hline
    Indian Premier League         & 6th edition of the Indian Premiere League cricket tournament. Tournament was amidst spot-fixing controversies during its last leg.                             \\
    Iran Earthquake               & Earthquake of magnitude 7.7 struck mountainous area in Iran, close to the border with Pakistan.                 \\
    Boston Blasts                 & Twin bomb blasts near the finish line during the final moments of the Boston Marathon. Police used Twitter to capture suspects.                                                          \\
    Kashmir Earthquake            & 5.8 magnitude earthquake struck Jammu and Kashmir, India.                                                                                                                                \\
    Mothers Day Shooting          &  20 killed in New Orleans, USA, when two armed men open fired during Mother's Day parade                                                                                                 \\
    Oklahoma Tornado              & Major tornado hit Oklahoma, invoking loss worth millions of dollars; impacting lives of thousands.                                                                                       \\
    London Terror Attack          & British Army soldier, Lee Rigby attacked and killed in a terrorist attack by Michael Adebolajo and Michael Adebowale in Woolwich, southeast London.                                      \\
    North India Floods            & Heavy rains, multiple burst clouds in the state of Uttarakhand, North India, causing major floods. Thousands dead, missing, and homeless.                                                \\
    Champions Trophy              & Last edition of the ICC Champions Trophy, a One Day International cricket tournament held in England and Wales.  \\
    Birth of Royal Baby           & Prince William and Kate Middleton of the British Royal Family gave birth to Prince George of Cambridge.                                                                                  \\
    Creation of Telangana         & Ruling party Indian National Congress resolved to request the Central government to form a separate state of Telangana, as the 29th state of India.                                      \\
    Washington Navy Yard Shooting & Lone gunman Aaron Alexis shot 12 people, injured 3 others in a mass shooting inside the Washington Navy Yard in Southeast Washington, D.C.                                               \\
    Cyclone Phailin               & Severe cyclonic storm Phailin was the second-strongest tropical cyclone ever to make landfall in India.                                                                                  \\
    Typhoon Haiyan                & Powerful tropical cyclone that devastated portions of Southeast Asia, particularly the Philippines.                                                        \\
    Metro North Train Derailment  & Metro-North Railroad Hudson Line passenger train derailed near the Spuyten Duyvil station in the New York City borough of the Bronx. 4 killed; 59 injured.                               \\
    Death of Nelson Mandela       & Nelson Mandela, the first President of South Africa elected in a fully representative democratic election, died at the age of 95 after suffering from a prolonged respiratory infection. \\ \hline
    \end{tabular}
\caption{Description of the 16 events for which we collected data.}
\label{tab:descofevents}
\vspace{-10pt}
\end{table*}

\subsection{Challenges with Facebook data}

Similar to Twitter, Facebook's Graph API provides keyword search capability, which can be used to search across all public historic data on Facebook. However, Facebook does not provide streaming capability for collection of relevant data in real time. Thus, in order to collect real time data from Facebook, we had to query the API continuously at regular intervals. The limited amount of data fields available from the Facebook API poses even bigger challenges with data analysis on Facebook. Most social networks have three fundamental types of data, viz. user profile, content, and network data. However, due to Facebook's highly restrictive privacy policies, the network data is virtually never available publicly. Also, the profile data provides nothing more than the first name, last name, and gender in most cases. User generated content is thus, the only part of data which is easily available publicly. The absence of network and profile data highly reduces the scope of a lot of analysis, which can be done on networks like Twitter. In this work, we are thus bound to limit our analysis on content.

\section{Content overlap} \label{sect:analysis}


This section presents our comparative analysis of Facebook and Twitter content. We also characterize Facebook content, and highlight the cross network content sharing habits of Facebook and Twitter users.

\subsection{Common textual content}

In an attempt to gauge the level of similarity between the content posted on Facebook and Twitter during an event, we extracted all the textual content from our complete dataset and looked for overlaps. To compare textual content, we extracted all unique keywords posted during each event on both social media, and stemmed them using Python NLTK's Porter Stemmer. Stemming is a natural language processing technique of reducing inflected words to their base, or root word. For example, stemming the words \emph{argumentative} and \emph{arguments} will both yield their root word \emph{argument}. We then compared Facebook and Twitter content for each event, and found an average overlap of 6.02\% ($\sigma$ = 2.3). The highest amount of overlap was found to be 9.7\% during the Indian Premiere League (IPL) cricket tournament. We noticed that despite the 140 character limit on tweets (and no limit on post length on Facebook), Twitter generated more unique stems during 13 out of the 16 events. On average, across the 16 events, Twitter generated 3.69 times the number of unique stems as Facebook ($\sigma$ = 4.9).~\footnote{We did not consider the Metro North Train Derailment event while calculating this value, since the data size was insignificant, and the Twitter:Facebook ratio for this event was skewed (161.03)} Interestingly, the 3 events which witnessed more unique stems on Facebook than on Twitter, happened to be totally local to India. These events were the Cyclone Phailin, the Floods in North India, and creation of the new state, Telangana. This hints towards more public activity on Facebook, than on Twitter during real world events occurring in India.

These overlap percentages, however, suffered from a drawback because of unequal content on Facebook and Twitter. For example, if Facebook generated $x$ unique stems, and Twitter generated $y = 4x$ unique stems during an event, mathematically, the overlap would not be more than 25\%, even if all $x$ unique stems on Facebook were present on Twitter. To dampen the effect of the disproportionate content sizes on the content overlap, we then calculated the content overlap in terms of the percentage of content on one social media which is also present on the other. This analysis revealed a much higher overlap. We discovered that for the 13 events where Twitter generated more content than Facebook, 30.56\% of content which was present on Facebook, was also present on Twitter ($\sigma$ = 17.4). Similarly, for the 3 events where Facebook generated more content than Twitter, 23.72\% of content which was present on Twitter, was also present on Facebook ($\sigma$ = 3.1). We believe that this is a significant amount of overlap in textual content on both these networks; especially considering the fact that the 140 character rate limit on Twitter introduces a lot of shortened words, and makes its content very different from the content posted on Facebook without a character limit.

We also compared content across events, for the three most popular events (in terms of the number of posts) on both Facebook and Twitter, viz. Boston Blasts, Oklahoma Tornado, and death of Nelson Mandela. Interestingly, we found a big overlap between Facebook content during Boston Blasts, and Oklahoma Tornado. Infact, Facebook content during Boston Blasts was more similar to Facebook content during the Oklahoma Tornado (24.07\% overlap), than Twitter content during Boston Blasts (14.12\% overlap). Similarly, Twitter content during Boston Blasts was found to be slightly more similar to Twitter content during the death of Nelson Mandela (8.89\% overlap), than Facebook content during the Boston Blasts (7.81\% overlap). This reflects the heterogeneous posting habits of Facebook and Twitter users, where content within one social network tends to be more similar during two different events, than content across two social networks during the same event.

To summarize, we found a significant overlap between Facebook and Twitter content during all the 16 events, and we also discovered that public Facebook content during two similar events tends to be more similar to each other, than Twitter content during the same events.

\subsection{Common hashtags}

Facebook in June 2013, rolled out the hashtag feature, which works similar to hashtags on Twitter~\cite{Facebook:2013}. We decided to study the hashtags posted on Facebook during the 16 events we studied, and compare them with the hashtags posted on Twitter. Table~\ref{tab:hashtags} shows the number of unique hashtags we found on Facebook and Twitter during the 16 events. On averge, Twitter produced 16.28 times the number of hashtags as Facebook ($\sigma$ = 29.8). Similar to content overlap, we calculated hashtag overlap on Facebook and Twitter, and found an average overlap of 6.02\% ($\sigma$ = 3.2) across the 16 events (max. 12.15\%, Cyclone Phailin). This value was almost identical to content overlap calculated in the previous section. Further, we found that for 14 out of the 16 events featuring more hashtags on Twitter than Facebook, 33.69\% hashtags present on Facebook, were also present on Twitter ($\sigma$ = 8.9). For the remaining 2 events, 19.08\% of hashtags present on Twitter, were also present on Facebook ($\sigma$ = 3.5). 

\begin{figure}[!ht]
     \begin{center}
\includegraphics[scale=0.18]{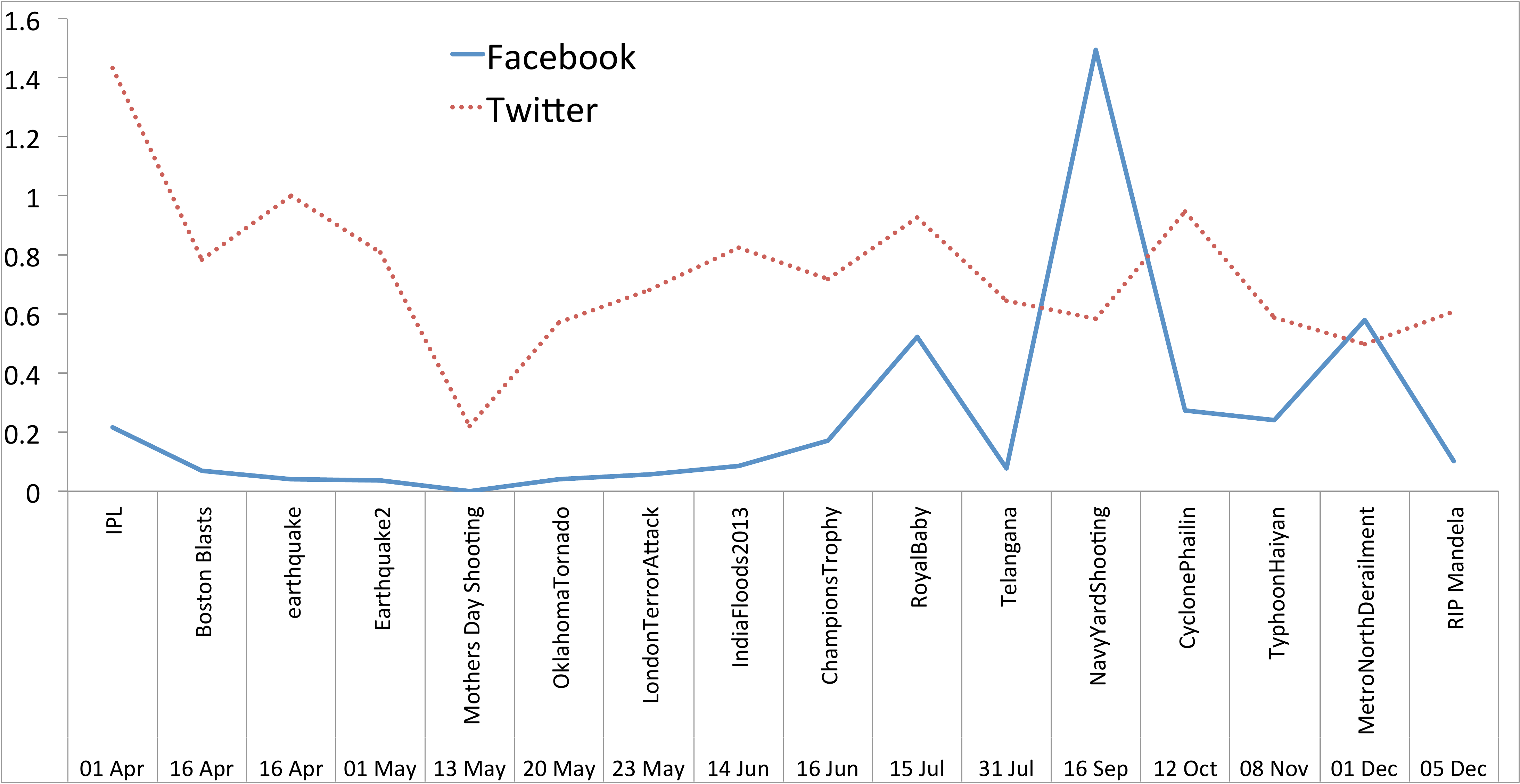}
    \end{center}
\vspace{-10pt}
    \caption{%
Number of hashtags per post on Facebook and Twitter. Hashtags have been present in Facebook posts even before Facebook launched official support in June, 2013.
     }
   \label{fig:hashtagsperpost}
\vspace{-7pt}
\end{figure}

As previously mentioned, Facebook officially launched hashtag support on June 12, 2013~\cite{Facebook:2013}, which means that 7 out of the 16 events in our dataset had already been concluded before this happened. This seemed reasonable explanation for the significantly low number of hashtags on Facebook, as compared to Twitter, in our dataset. However, we found notable presence of hashtags even before Facebook officially rolled out hashtag support. Figure~\ref{fig:hashtagsperpost} shows a graph of the number of total hashtags found per post, during the 16 events on Facebook and Twitter. The events are arranged in chronological order from left to right. As is evident from the graph, hashtags were posted on Facebook even before June 12. However, a notable increase in the number of hashtags per Facebook post is visible after the India Floods event, for which, we started data collection on June 14. One possible reason for the presence of hashtags on Facebook even before the official launch of its support, could have been Twitter users pushing their tweets to Facebook via apps like HootSuite, Facebook for Twitter, etc. However, this wasn't true, since we found that over 90\% of Facebook posts which contained hashtags, came from Facebook (web and mobile apps) itself. It is also evident that the number of hashtags used on Twitter is much more than that on Facebook, during almost all events.

Twitter was the first social media service to start the use of hashtags in 2009.~\footnote{\url{http://twitter.about.com/od/Twitter-Hashtags/a/The-History-Of-Hashtags.htm}} The presence of hashtags on Facebook even before official support, thus highlights how users tend to carry their posting habits across multiple social networks; in this case, from Twitter to Facebook. The statistics we presented in this subsection, along with those from the previous subsection, indicate that the public content on Facebook and Twitter during real world events is reasonably similar in terms of text, and hashtags. However, the amount of public content produced by Twitter is much more than that produced by Facebook during real world events. To the best of our knowledge, no work in the past has studied any aspects of the use of hashtags on Facebook.

\begin{table}
\small
\begin{center}
    \begin{tabular}{l|l|l|l}
    \hline
    Event                        & Facebook & Twitter & Common \\ \hline
    North India Floods           & 5,867     & 14,127   & 1,296   \\
    Boston Blasts                & 20,538    & 74,889   & 6,222   \\
    Oklahoma Tornado             & 14,487    & 56,541   & 4,844   \\
    Mothers Day Shooting         & 0        & 496     & 0      \\
    London Terror Attack         & 1,558     & 10,648   & 678    \\
    IPL       & 12,438    & 66,702   & 4,434   \\
    Navy Yard Shooting & 910      & 9,669    & 465    \\
    Champions Trophy             & 9,925     & 65,371   & 3,044   \\
    Birth of Royal Baby          & 8,504     & 72,985   & 3,862   \\
    Kashmir Earthquake           & 35       & 3,609    & 11     \\
    Iran Earthquake              & 81       & 6,402    & 16     \\
    Creation of Telangana        & 6,378     & 5,781    & 1,306   \\
    Cyclone Phailin              & 2,898     & 4,143    & 763    \\
    Typhoon Haiyan               & 36,256    & 24,793   & 3,862   \\
    Metro North Derailment  & 182      & 1,474    & 75     \\
    RIP Nelson Mandela      & 29,787    & 89,193   & 8,096   \\ \hline
    \end{tabular}
\end{center}
\vspace{-10pt}
\caption{Number of unique hashtags posted during the 16 events on Facebook and Twitter.}
\label{tab:hashtags}
\vspace{-12pt}
\end{table}

\subsection{Timeliness of content}



For unprecedented events where timeliness can be critical, we extracted the timestamps of the first Facebook post published about the event. We could not extract the correct first tweet about any such event, since unlike Facebook, Twitter's search API is not exhaustive, and does not return all past tweets. Also, given that such events are unexpected, it was hard to initiate data collection instantly after an event took place, and be able to collect the very first tweet. We sorted the posts in our dataset, in increasing order of their publishing time, and manually read the post content to find the first appearance of a relevant post. We observed that for the 6 such events that we investigated (viz. Boston Blasts, London Terror Attack, Washington Navy Yard Shooting, Kashmir Earthquake, Iran Earthquake, and Metro North Train Derailment), the first post on Facebook appeared, on average, 11 minutes 1.5 seconds ($\sigma$ = 10 minutes 11 seconds) after the event took place in the real world (Max. 25 minutes, 53 seconds, Washington Navy yard Shooting). Interestingly, during the Boston Marathon Blasts, the first Facebook post occurred just 1 minute 13 seconds after the first blast, which was 2 minutes 44 seconds before the first tweet~\cite{gupta20131}.

Osborne et al. conducted a similar experiment, and reported a much higher latency of 2.36 hours for Twitter, and 9.89 hours for Facebook over 29 events~\cite{Osborne:2014}. One possible reason for this higher latency could be the difference in the kind of events we tracked. While Osborne et al. identified events from a Twitter stream of trending topics, we focused our analysis on unprecedented events, which do not become trending topics immediately after taking place in the real world.

This analysis depicts that when an unprecedented event takes place in the real world, it appears very quickly on Facebook. With results from our previous sections, we conjecture that Facebook can be a rich and timely source of information during real world events. However, one rare question which remains unexplored is, if there exists content which appears on both Twitter and Facebook at the same time. We explore the same in the next subsection.

\subsection{Content appearing at the same time}

\begin{figure}[!ht]
     \begin{center}
        \subfigure[URL sources on Facebook]{%
           \label{fig:fb_nosp}
           \includegraphics[scale=0.34]{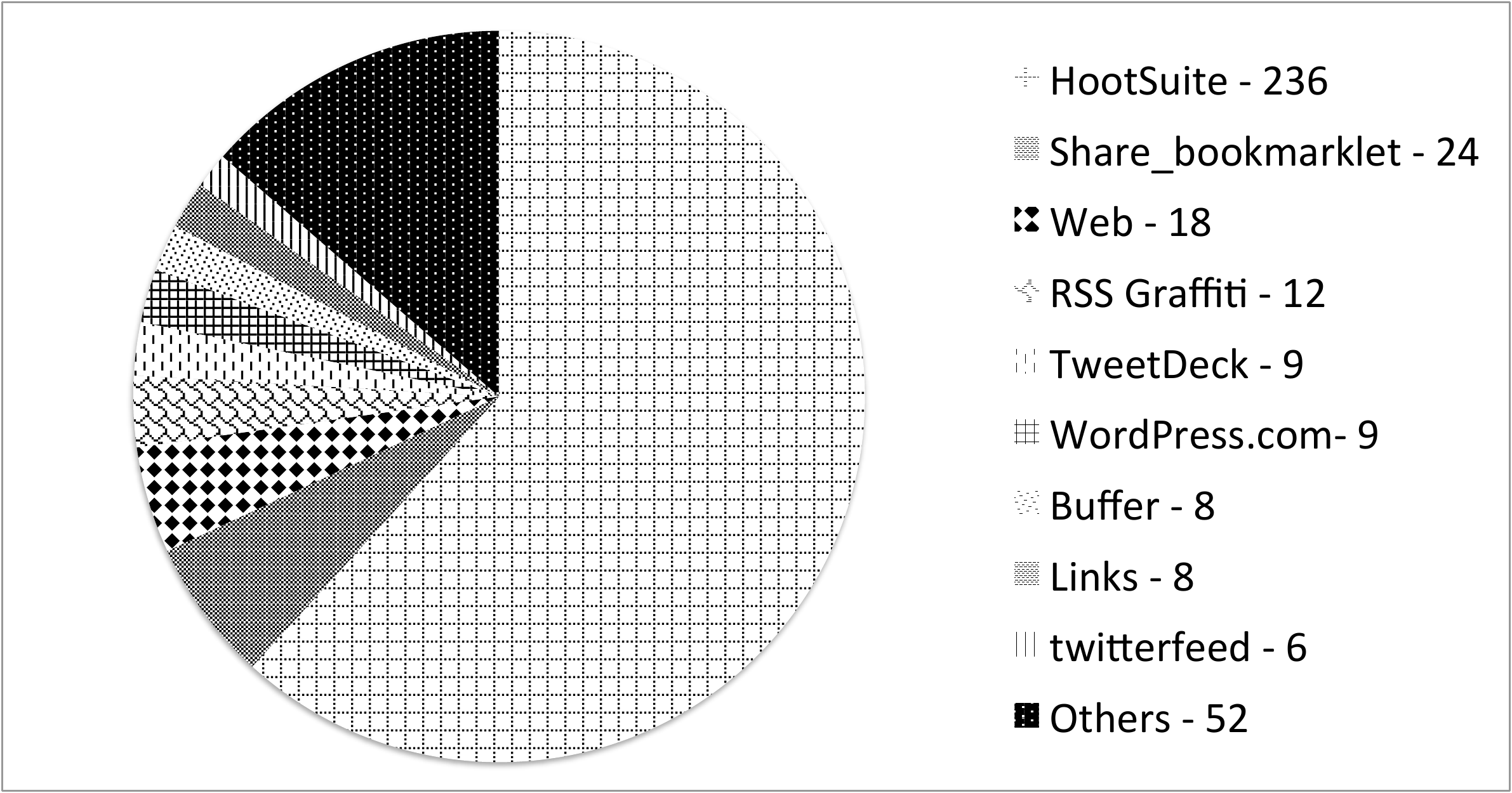}
        }
        	\subfigure[URL sources on Twitter]{%
            \label{fig:tw_nosp}
            \includegraphics[scale=0.34]{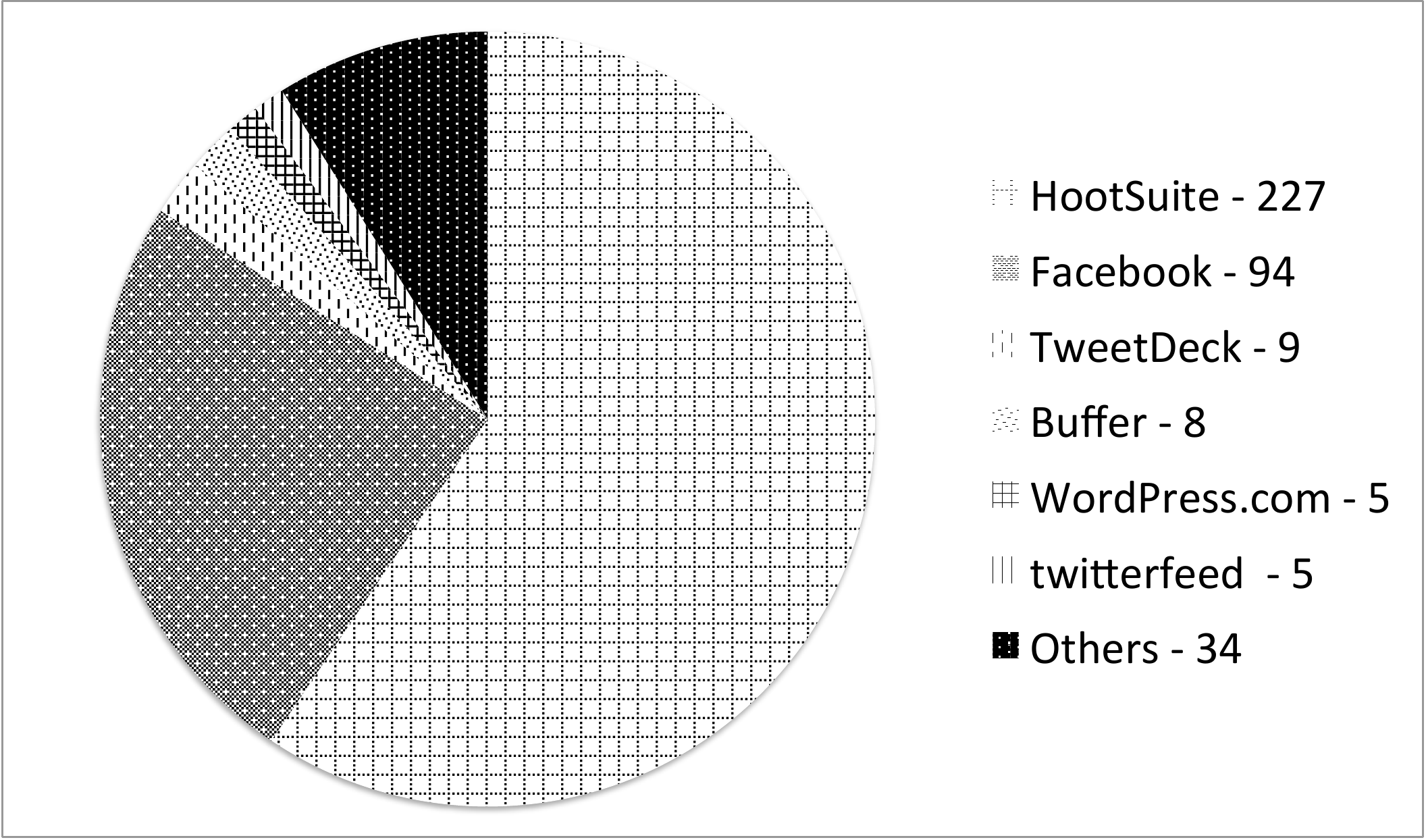}
        }
    \end{center}
\vspace{-10pt}
    \caption{%
URL sources on Facebook and Twitter for the 382 URLs which appeared on both the networks at the same time. A majority of these URLs were shared using famous cross content sharing platforms like HootSuite, TweetDeck, Twitterfeed etc.
     }
   \label{fig:sametimeurlsources}
\end{figure}

We extracted a total of 57,227 URLs which appeared on both Facebook and Twitter during the 16 events, and found that 382 of these URLs were posted on both the networks at exactly the same time. We extracted the app / platform using which these URLs were posted, and found that 271 out of the 382 URLs (70.94\%) were posted using the same source / social media management platform on both the networks. For the remaining 111 URLs, although the app names did not match, we discovered that they were also posted using cross platform content sharing sources, which connect users' Facebook and Twitter accounts, like the Twitter app for Facebook etc. Figure~\ref{fig:sametimeurlsources} represents the distribution of the various platforms used for all the 382 URLs on both networks. HootSuite clearly appeared to be the most commonly used platform for posting content across the two networks.

With such observations, it appeared that these URLs may have been posted from Facebook and Twitter accounts belonging to the same real world user. To confirm our findings, we extracted the user handles, and user names of all these users on both Facebook, and Twitter, and computed a similarity score using the Jaro Distance metric~\cite{jaro1978unimatch}. Four similarity scores were computed for each URL:

\begin{enumerate}
\item $Twitter_{handle}$ : $Facebook_{handle}$
\vspace{-3pt}
\item $Twitter_{handle}$ : $Facebook_{name}$
\vspace{-3pt}
\item $Twitter_{name}$ : $Facebook_{handle}$
\vspace{-3pt}
\item $Twitter_{name}$ : $Facebook_{name}$
\end{enumerate}

We found that for 265 out of the 382 URLs (69.37\%), at least one of the 4 aforementioned scores came out to be greater than 0.85. In fact, for 209 URLs (54.71\%), at least one of these 4 values came out to be 1, i.e. exact match. We could not determine Facebook usernames and handles for 19 users; these accounts were either deleted, or the username and handle fields were not available. With this analysis, it is safe to conclude that at least 70\% of URLs which are posted on two social networks (Facebook and Twitter in this case) at the same time, are posted by the same real world user. Like URLs, this analysis can also be performed by looking at timestamps of posts containing identical textual content on Facebook and Twitter, and could possibly yield an even bigger set of users who post content on both these networks at the same time. This characteristic can be a helpful feature in connecting two online identities across social networks to a single real world person, where existing accuracy rates have much scope of improvement~\cite{jain2013seek}. However, performing this analysis is beyond the scope of our work.


\subsection{Inter-network URLs}

Apart from studying common URLs which appeared on both networks, we investigated the amount of inter-network content sharing on Facebook and Twitter. We extracted and fully expanded all the URLs shared across all 16 events on both these networks. In total, we got a total of 2.11 million URLs posted on Facebook, and 6.90 million URLs posted on Twitter. We then extracted the domains of all these URLs, and found that Twitter shared more \emph{facebook.com} URLs than the number of \emph{twitter.com} URLs shared on Facebook. On average, 2.5\% ($\sigma$ = 2.1) of all URLs shared on Twitter belonged to the \emph{facebook.com} domain, but only 0.8\% ($\sigma$ = 1.0) of all URLs shared on Facebook belonged to the \emph{twitter.com} domain. On average (during the 16 events), in terms of ranking, \emph{twitter.com} was the 17$^{th}$ most shared domain on Facebook ($\sigma$ = 9.3), and \emph{facebook.com} was the 12$^{th}$ most shared domain in Twitter ($\sigma$ = 12.5). Further, in terms of percentage of URLs, Twitter URLs were most popular on Facebook during the Navy Yard Shooting event; while Facebook URLs were most popular on Twitter during the North India Floods.

Table~\ref{tab:crossurls} presents the ranking and percentage of Twitter, YouTube, and Instagram URLs shared on Facebook; and Facebook, YouTube and Instagram URLs shared on Twitter. Interestingly, we found YouTube to be ranked very high on both Facebook and Twitter during most events. On average, YouTube was ranked 3$^{rd}$ ($\sigma$ = 1.4) and 7$^{th}$ ($\sigma$ = 4.4) on Facebook and Twitter respectively. Instagram was ranked fairly low on Facebook (amongst top 100 during only 4 out of the 16 events); but was amongst the top 100 during 14 out of the 16 events on Twitter.

\begin{table*}[!ht]
\begin{center}
    \begin{tabular}{l|c|c|c|c|c|c|c|c|c|c|c|c}
    \hline
    ~                      & \multicolumn{6}{c|}{Facebook} & \multicolumn{6}{c}{Twitter} \\ \cline{2-13}
    ~                      & \multicolumn{2}{c|}{Twitter} & \multicolumn{2}{c|}{YouTube} & \multicolumn{2}{c|}{Instagram} & \multicolumn{2}{c|}{Facebook} & \multicolumn{2}{c|}{YouTube} & \multicolumn{2}{c}{Instagram} \\ \cline{2-13}
    Event                      & Rank     & \% & Rank    & \% & Rank      & \% & Rank     & \% & Rank    & \% & Rank      & \% \\ \hline
    North India Floods     & 23       & 0.15    & 3       & 4.0     & 324       & 0.006   & 3        & 7.18    & 6       & 4.50    & 7         & 4.49    \\
    Boston Blasts          & 5        & 1.91    & 2       & 2.65    & 172       & 0.06    & 8        & 1.28    & 2       & 3.97    & 4         & 2.00    \\
    Oklahoma Tornado       & 17       & 0.34    & 2       & 6.24    & 107       & 0.05    & 2        & 5.38    & 3       & 3.77    & 5         & 2.63    \\
    Mothers Day Shooting   & -        & -       & 4       & 25      & -         & -       & 31       & 0.46    & 8       & 2.72    & 199       & 0.05    \\
    London Terror Attack   & 16       & 0.63    & 2       & 9.34    & 409       & 0.01    & 15       & 0.93    & 5       & 5.60    & 58        & 0.14    \\
    IPL                    & 14       & 0.59    & 2       & 5.36    & 247       & 0.01    & 4        & 4.03    & 6       & 3.78    & 28        & 0.54    \\
    Navy Yard Shooting     & 3        & 4.32    & 5       & 2.94    & 40        & 0.35    & 22       & 0.70    & 5       & 2.91    & 97        & 0.13    \\
    Champions Trophy       & 27       & 0.15    & 2       & 4.82    & 192       & 0.02    & 2        & 5.42    & 3       & 5.29    & 5         & 3.85    \\
    Birth of Royal Baby    & 10       & 1.17    & 2       & 3.89    & 52        & 0.16    & 11       & 1.19    & 8       & 1.38    & 6         & 1.74    \\
    Kashmir Earthquake     & -        & -       & -       & -       & -         & -       & 12       & 0.89    & 11      & 1.45    & 39        & 0.22    \\
    Iran Earthquake        & 34       & 0.49    & 1       & 24.50   & -         & -       & 16       & 1.13    & 10      & 2.66    & 29        & 0.50    \\
    Creation of Telangana  & 33       & 0.03    & 2       & 10.74   & 717       & 0.001   & 6        & 4.88    & 3       & 5.71    & 32        & 0.36    \\
    Cyclone Phailin        & 17       & 0.90    & 5       & 1.81    & 662       & 0.009   & 5        & 2.48    & 11      & 1.16    & 123       & 0.09    \\
    Typhoon Haiyan         & 16       & 0.53    & 2       & 4.81    & 20        & 0.50    & 4        & 2.75    & 9       & 1.90    & 12        & 1.45    \\
    Metro North Derailment & 12       & 1.28    & 6       & 2.56    & 48        & 0.32    & 48       & 0.34    & 19      & 1.11    & 34        & 0.57    \\
    RIP Nelson Mandela     & 12       & 0.29    & 2       & 8.32    & 134       & 0.03    & 4        & 1.95    & 3       & 3.17    & 5         & 1.82    \\ \hline
    \end{tabular}
\vspace{-10pt}
\caption{Rank and percentage of Twitter / Facebook, YouTube, and Instagram URLs on Facebook / Twitter. Facebook URLs were found to be more popular on Twitter, than Twitter URLs were popular on Facebook. YouTube was found to be one of the most famous domains on both Facebook and Twitter during almost all events.}
\label{tab:crossurls}
\end{center}
\vspace{-15pt}
\end{table*}

These numbers highlight that apart from posting similar content during real world events, Facebook and Twitter users also post a significant amount of content picked from each other. It was also interesting to see that YouTube was one of the most famous domains shared on both the networks. It could be interesting to perform a similar content overlap analysis for Facebook, Twitter, and YouTube, and see how much YouTube content is present on both these networks.

\section{Qualitative analysis}  \label{sect:qual}

To get a qualitative overview of the content, and extract a true positive dataset of spam, we decided to extract the \emph{most active users}, and their posts from both Facebook and Twitter as a sample. Ideally, analyzing the entire dataset would have been a good approach to compare the spam content of both the social networks during the 16 events. However, such amount of data would have been extremely difficult to annotate manually, and extract true positive spam. We thus decided to analyze content posted by only the most active users during these events, to get a qualitative overview of the content, and for our spam analysis. The reason for choosing this approach over random sampling was two-fold. Firstly, since the users we pick with this approach contribute the maximum content, the probability of their post appearing in a search executed by a common user would be more than that of a post made by a random user. We believe that it would be interesting to study the content and users that a common user comes across, and consumes from social media during a real-world event. Statistically, pieces of content generated by the most active users would have a higher probability of popping up during searches made by common users, and hence useful to study. Secondly, normal random sampling approaches would have led to a biased overall dataset in terms of numbers, because of the varying number of users and posts generated during different events. For example, a 1\% random sample would have generated over 12,000 Facebook users during IPL, but only 10 users during the Kashmir earthquake (Table~\ref{stats}). Our approach of picking the most active users is scalable, and yields a comparable and manageable number of users for almost all events, irrespective of the total number of users or posts generated during an event.

\subsection{Selecting most active users} \label{sec_extract_mau}

To extract the most active users, we calculated the \emph{content gain} factor associated with each user present in our dataset. The \emph{content gain} factor associated with a user was calculated by the fraction of content added by the user during an event through the number of posts he / she made during the event. For each event, we sorted the users in decreasing order of the number of posts made by them during the event. Starting from the ``most active" user in terms of the number of posts, we calculated the content gain as follows:

\begin{algorithm}
c\_sum = 0
\begin{algorithmic}
\ForAll{users}
\\
c\_sum = c\_sum + num\_posts\_by\_user
    \If {(num\_posts\_by\_user / c\_sum) $\times$ 100 $>$ $k$}
        \State Consider user
    \Else
        \State Ignore user
    \EndIf
\EndFor

\label{algo1}
\caption{Extracting most active users}
\end{algorithmic}
\end{algorithm}

In Algorithm 1, $k$ is a tunable parameter which can be chosen as required. The value of $k$ is directly proportional to the content generated by a user chosen; thus, a high value of $k$ would only yield very highly active users. For our experiment, we picked a value of $k$ = 3, which left us with a total of 252 unique Twitter users contributing 222,529 tweets, and 220 unique Facebook users contributing 66,688 posts, across the 16 events. Choosing a lower value of $k$ could have increased these numbers, and made our sample richer and more representative, but at the same time, it would have made the sample very hard and unmanageable to annotate and analyze manually.

\subsection{Manual classification of most active users} \label{sect:mau}

\begin{figure}[!ht]
     \begin{center}
        \subfigure[Spam on Facebook]{%
           \label{fig:fb_spam}
           \includegraphics[scale=0.3]{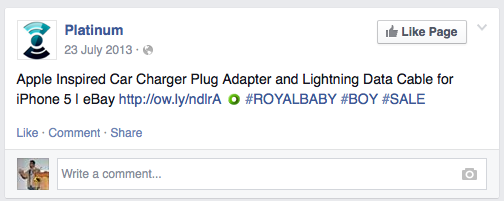}
        }
        	\subfigure[Spam on Twitter]{%
            \label{fig:tw_spam}
            \includegraphics[scale=0.23]{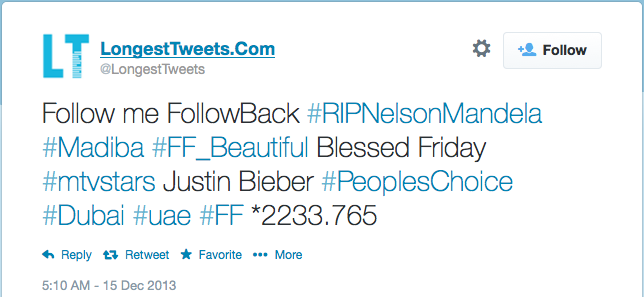}
        }
    \end{center}
\vspace{-10pt}
    \caption{%
Examples of spam posts on Facebook and Twitter. 
     }
   \label{fig:spamexample}
\vspace{-10pt}
\end{figure}

We manually went through the user profiles and content posted by the 472 most active users (252 from Twitter + 220 from Facebook) which we extracted from the previous step, and marked these users as spammers, and non-spammers. We did not use multiple annotators, or inter-annotator agreement strategies for this annotation, as our annotation classes were distinct and non-confusing. Spammers were marked so because of highly repetitive and / or irrelevant content. Figure~\ref{fig:spamexample} shows an example of a spam post on Facebook and Twitter. Table~\ref{tab:mau} represents the detailed statistics of the users in our most active users dataset.

\begin{table}[!h]
\begin{center}
    \begin{tabular}{l|l|l}
    \hline
    Category            & Facebook & Twitter \\ \hline
    Users               & 48       & 242     \\
    Pages               & 160      & -       \\
    - Verified            & 7        & 7       \\
    - Spammers            & 25       & 37      \\
    - False positives     & 29       & 33      \\ 
    Suspended           & -        & 7       \\
    Deleted / Not found & 12        & 3       \\ \hline
    Total & 220 & 252 \\ \hline
    \end{tabular}
\end{center}
\vspace{-10pt}
\caption{Detailed statistics of the most active users on Facebook and Twitter captured during 16 real-world events.}
\label{tab:mau}
\vspace{-10pt}
\end{table}

\subsection{Qualitative overview} \label{sec:overview}
After removing the false positives, we were left with a total of 191 Facebook users, contributing 59,148 posts, and 219 Twitter users contributing 196,880 tweets.~\footnote{False positives were users whose content was captured merely because their usernames matched at least one event related keyword.} Surprisingly, Facebook content posted by the most active users was found to be highly irrelevant to the events during which, the data was collected. Infact, the content looked alarmingly polar, and radical. This spam and propaganda content completely overshadowed the relevant content about the events under consideration, highlighting the prevailing nature of \emph{topic hijacking} on Facebook, which has been observed previously on Internet forums. The tag cloud of the most frequently occurring words in this content looked very similar to the tag cloud of spam content, shown in Figure~\ref{fig:fb_spam}, and did not give a clear picture. Upon further manual inspection, we found that this behavior was due to a few spammers who were posting enormously large pieces of such propaganda content repeatedly, during the Royal Baby event; thus overshadowing all the normal length relevant posts. On the other hand, Twitter content looked fairly usual, and most of the top keywords were related to one of the 16 events. 

We then removed the spammers in addition to false positives, and repeated this analysis for both Facebook and Twitter to get a clearer picture of the relevant content posted by the most active users on both these networks. Removing spammers in addition to false positives yielded a total of 169 Facebook users (contributing 52,615 posts), and 182 Twitter users (contributing 173,325 tweets). Figures~\ref{fig:fb_nosp} and~\ref{fig:tw_nosp} represent the top 100 most frequently occurring keywords in this content. 

\begin{figure}[!ht]
     \begin{center}
        \subfigure[Facebook content without spam and false positives]{%
           \label{fig:fb_nosp}
           \includegraphics[scale=0.34]{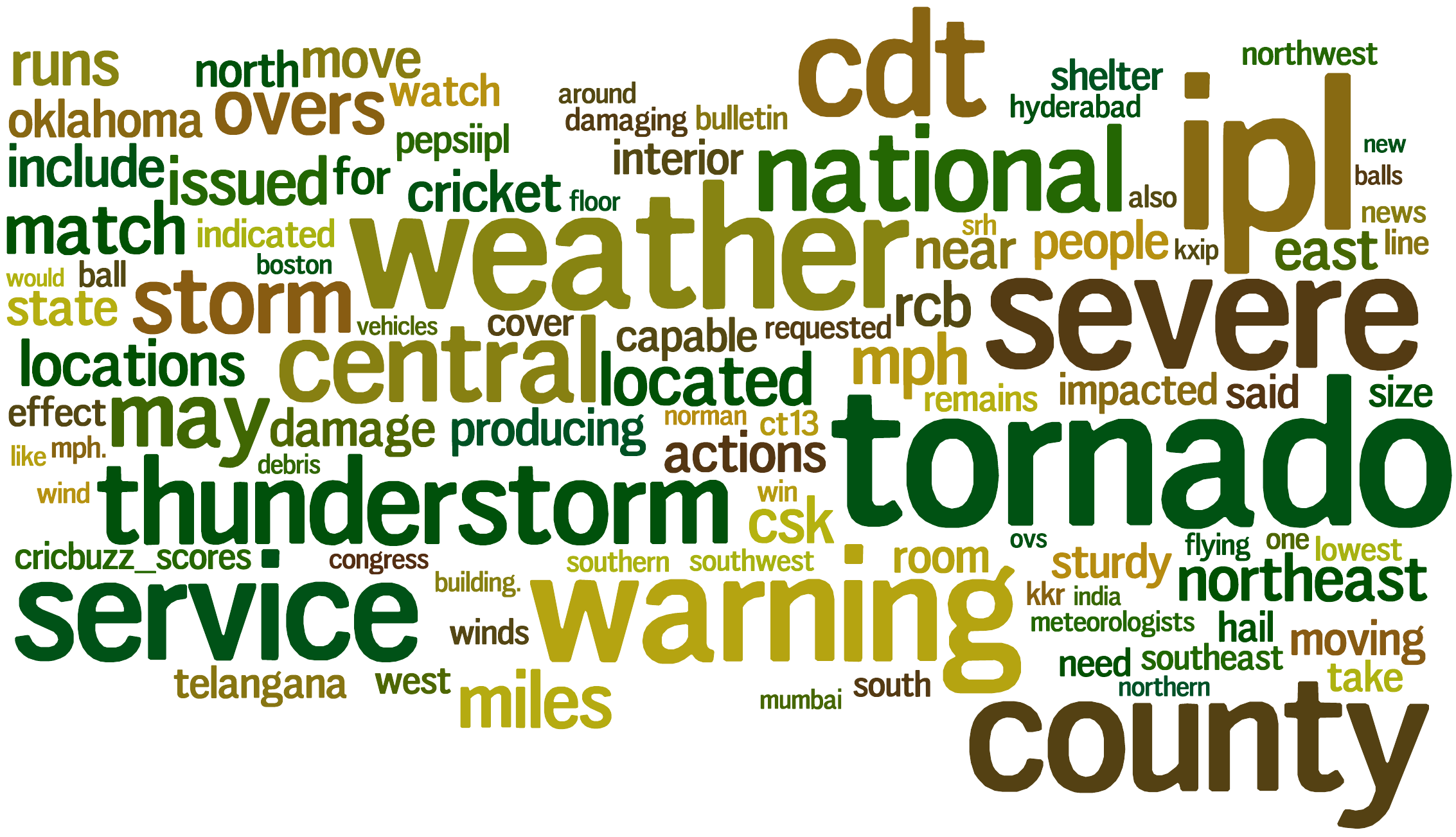}
        }
        	\subfigure[Twitter content without spam and false positives]{%
            \label{fig:tw_nosp}
            \includegraphics[scale=0.35]{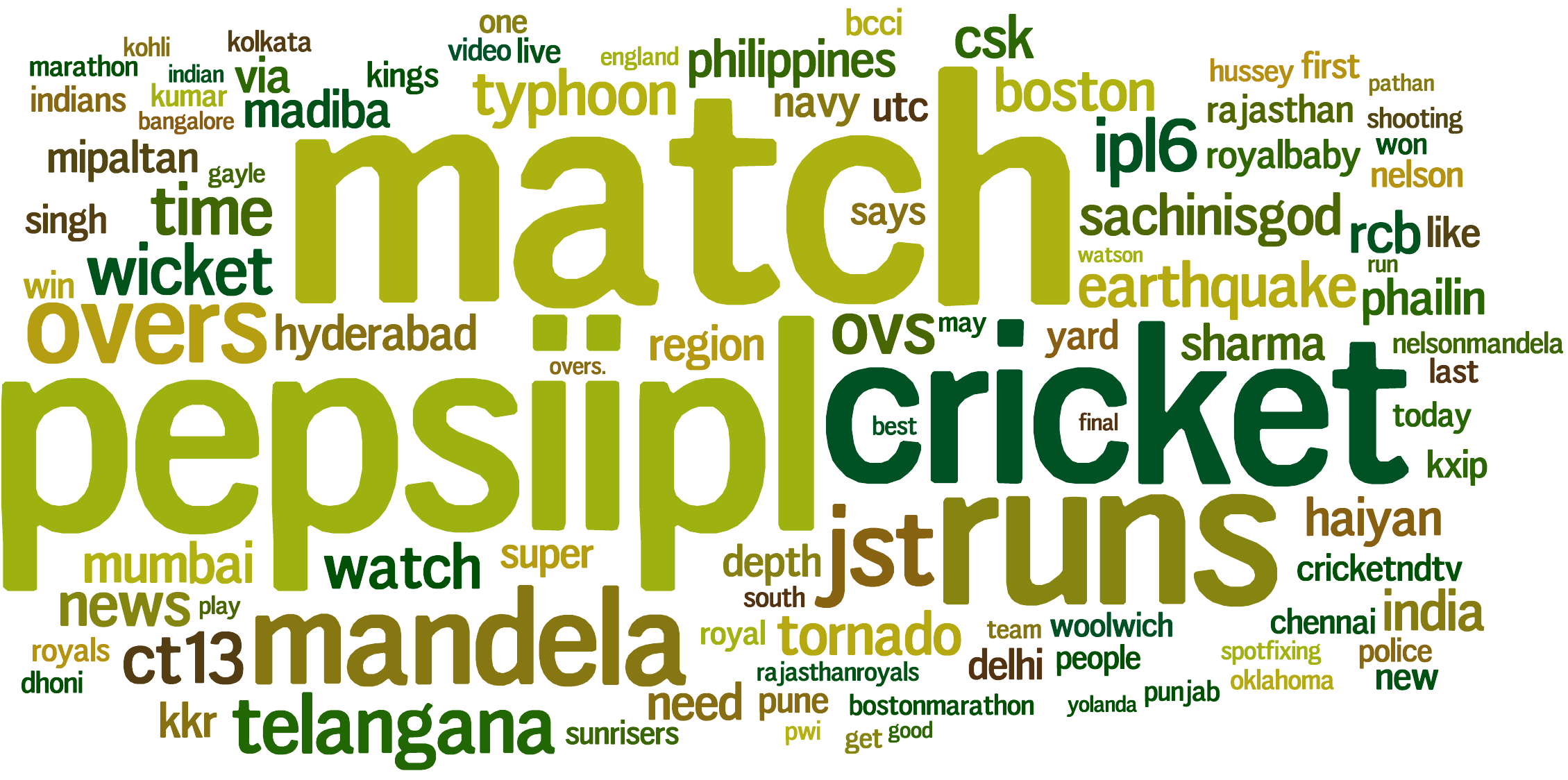}
        }
    \end{center}
\vspace{-15pt}
    \caption{%
Tag cloud of the 100 most frequently posted words by most active users on Facebook and Twitter during 16 real-world events. Facebook content looks totally irrelevant to the events.
     }%
   \label{tag_cloud_mau}
\vspace{-12pt}
\end{figure}

The alarmingly polar content overshadowing relevant content on Facebook prompted us to further investigate spam posted by the most active users on Facebook, and compare it with spam on Twitter. We discuss our in-depth analysis on spam in Section~\ref{sect:spam}.

\section{Spam analysis} \label{sect:spam}

As discussed in Section~\ref{sec:overview}, studying Facebook content revealed an alarming amount of spam content and users, which we decided to investigate in detail. We observed that contrary to regular behavior, where more than 75\% content was generated by Facebook pages (Table~\ref{tab:mau}); in case of spam, 68\% contributors were users. In a recent article,  Facebook also pointed that a lot of spam posts are published by pages; and announced improvements to reduce such spam in their \emph{News Feed}.~\footnote{\url{http://newsroom.fb.com/news/2014/04/news-feed-fyi-cleaning-up-news-feed-spam/}}

In terms of events, we observed that spam was present in only 5 out of the 16 events on Facebook, as opposed to Twitter, where spam was found to be present in 13 out of the 16 events. We discovered a striking difference between spam on Facebook and Twitter as well. Figures~\ref{fig:fb_spam} and~\ref{fig:tw_spam} represent the tag cloud of the top 100 most frequently occurring words in spam posts on Facebook and Twitter respectively. As is evident from the figures, spam content on Facebook looked completely non-relevant to the events under consideration, while spam content on Twitter was a mix of event related keywords and other famous keywords, viz. Justin Bieber, Apple, Obama etc., which are usually trending on Twitter, and on the web in general. 

\begin{figure}[!ht]
     \begin{center}
        \subfigure[Spam content on Facebook]{%
           \label{fig:fb_spam}
           \includegraphics[scale=0.34]{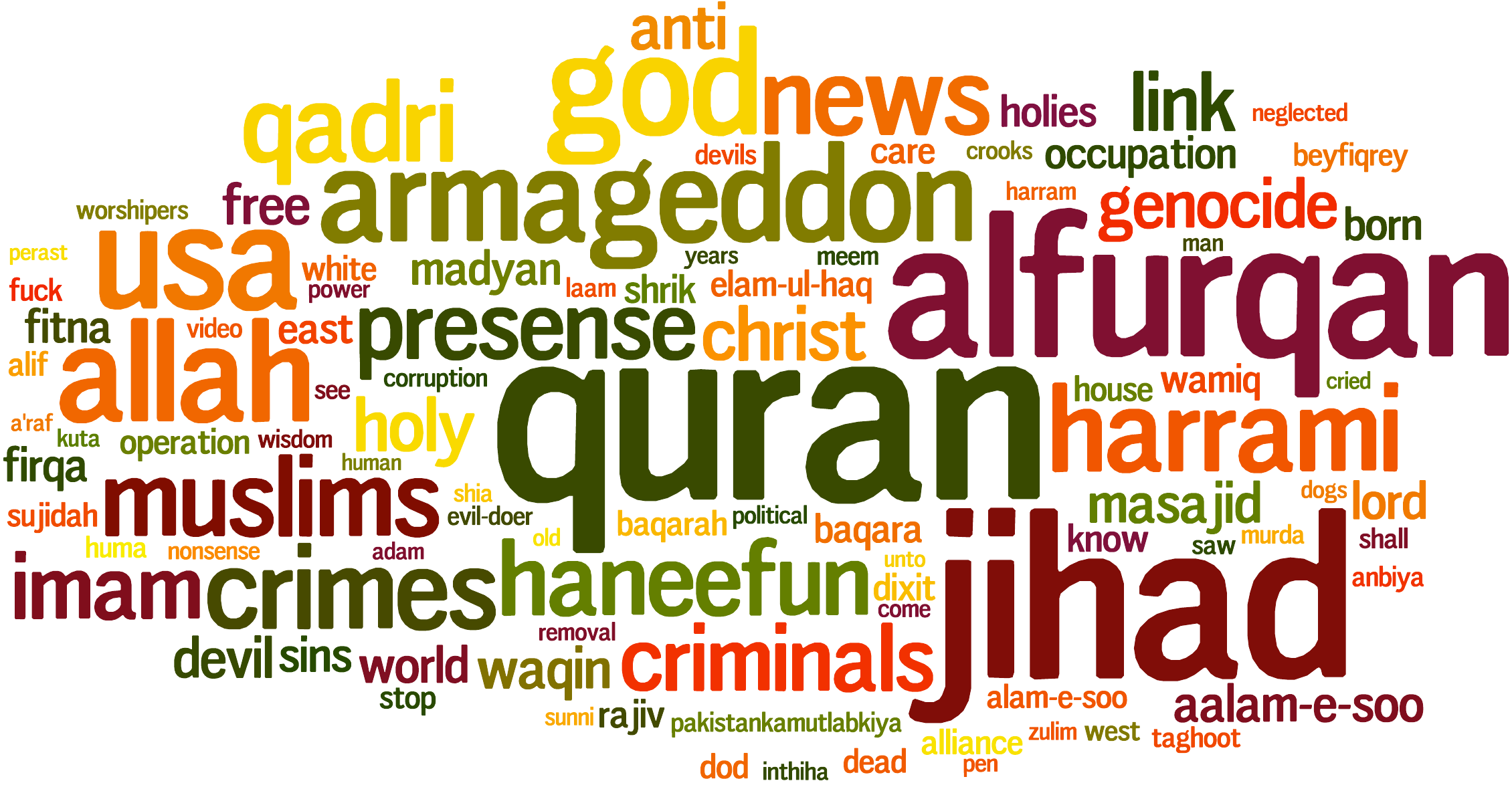}
        }
        	\subfigure[Spam content on Twitter]{%
            \label{fig:tw_spam}
            \includegraphics[scale=0.35]{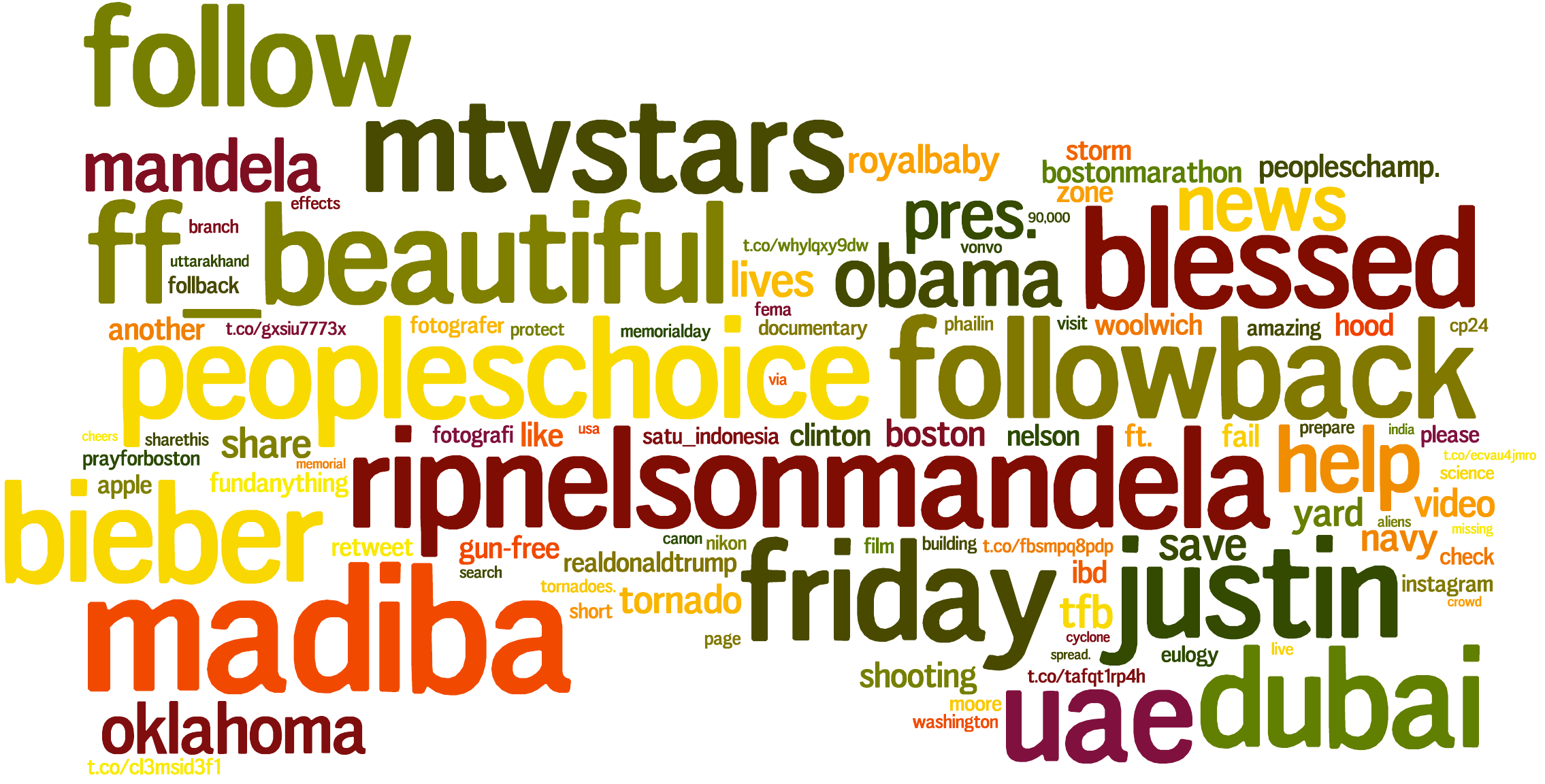}
        }
    \end{center}
\vspace{-15pt}
    \caption{%
Spam content on Facebook and Twitter. Facebook spam was found to be radical in nature, irrelevant to the events under consideration, and very different from Twitter spam.
     }
   \label{fig:spam}
\vspace{-12pt}
\end{figure}

\subsection{Facebook spam classification}
This phenomenal difference between spam content on Facebook and Twitter prompted us to probe further. Apart from discovering that more users post spam than pages, we noticed some more distinct characteristics about spam content on Facebook. These included enormously large pieces of content, intense repetition, more frequent presence of URLs, etc. in spam posts. We decided to extract these features and apply machine learning algorithms on our labeled dataset of spam and non-spam posts that we extracted from our manually annotated dataset of most active users (Section~\ref{sect:mau}). There have been multiple attempts to detect spam on Twitter in the past~\cite{Benevenuto2010,grier2010spam,mccord2011spam,wang2010don}. However, to the best of our knowledge, this is one of the first attempts to detect spam posts on Facebook using machine learning techniques. Conforming to our preliminary observations about notable differences between spammers and non-spammers, we were able to achieve a maximum accuracy of 99.9\% using the Random Forest Classifier trained on a set of 21 features, as listed in Table~\ref{tab:feats}.

\begin{table}[!h]
\small
    \begin{tabular}{p{2.7cm}|p{5cm}}
    \hline
    Feature               & Description                                         \\ \hline
    post\_richness$^*$ & $\frac{post\_words}{post\_chars}$ \\
    post\_length & Total length of the post                            \\
    post\_chars & Number of characters in the post                    \\
    post\_rep\_factor & $\frac{post\_unique\_words}{post\_words}$ \\
    post\_words & Number of  words in the post                        \\
    post\_unique\_words & Number of unique words in post                  \\
    isPage                & True if the post comes from a page; False otherwise \\
    type                  & Status / Link / Photo / Video                       \\
    num\_fb\_urls  & Number of Facebook.com URLs                          \\
    num\_urls$^*$   & Number of URLs in the post                          \\
    app                   & Application used to post                            \\
    num\_likes  & Number of likes on the post                         \\
    pageLikes             & Number of likes on the page if isPage is True       \\
    category              & Category of page if isPage is True              \\
    num\_hashtags$^*$ & Number of hashtags in the post                      \\
    num\_unique\_hashtags & Number of unique hashtags in the post               \\
    num\_shares & Number of times the post has been shared            \\
    hashtag\_rep\_factor & $\frac{num\_unique\_hashtags}{num\_hashtags}$ \\
    app\_ns     & Namespace of application               \\
    num\_short\_urls & Number of short URLs in the post                \\
    num\_comments & Number of comments on the post                      \\

 \hline
    \end{tabular}
\caption{List of features used for spam classification on Facebook, in decreasing order of significance.}
\label{tab:feats}
\vspace{-15pt}
\end{table}

Some of the features marked with $^*$ in Table~\ref{tab:feats} have been previously used for spam classification on Twitter~\cite{Benevenuto2010} and emails~\cite{Toolan2010}. We introduced two new stylometric features viz. \emph{post repetition factor}, and \emph{hashtag repetition factor} to capture the amount of repetition within a post, which haven't been used in the past for spam classification. These features are a numeric value ranging from 0 to 1, where a low value signifies higher repetition. Most of the other features like number of likes, number of comments, number of shares, application name, application namespace, category, type, pageLikes etc. can be directly obtained from Facebook's Graph API.

We ran the Naive Bayesian, J48 Decision Tree, and Random Forest classifiers on our labeled dataset of 7,882 spam posts, and 58,806 non-spam posts from the most active users, using the aforementioned set of 21 features. The Random Forest classifier achieved the maximum accuracy of 99.90\%. This exceptionally high value was not surprising, as we were able to find notable differences in spam and non-spam posts during our initial manual investigation itself. Researchers in the past have been able to achieve over 95\% accuracy for classifying spam on Twitter too~\cite{mccord2011spam}, but these accuracy rates correspond to a general stream of data, and with an extremely rich feature set. In contrast, the dataset we used for our spam classification contained posts specific to real world events, and coming from only the most active users. This makes the task of spam identification easier for the classification algorithms, since the input data to the classifier is highly restricted. Note that these results may differ drastically on a general, random sample of Facebook posts. Our objective is not to propose new, or compete with existing feature sets and machine learning algorithms for detecting spam on Facebook and Twitter, but to highlight the fact that spam posted during real world events by the most active users differs drastically from genuine content.

From our classification results, we noticed that all the 6 post level features viz. post richness, post length, number of words, number of unique words, number of characters, and post repetition factor came out to be the highest ranked (most significant) features. We then decided to remove these 6 features, and run the algorithms again on a depleted feature set containing only 15 out of the 21 features as mentioned in Table~\ref{tab:feats}. Surprisingly, we were still able to achieve an exceptionally high accuracy of 99.30\% using the Random Forest classifier. It was interesting to see that the number of likes, comments, and shares did not turn out to be important features. These results also bring out the fact that it might be easier to detect spam on Facebook, as compared to what we have learnt about Twitter from existing literature. However, we would like to extend our analysis on a bigger labeled dataset, and verify our results in future. In order to do so, we intend to generate a larger manually annotated labeled dataset for Facebook spam using multiple human annotators. 

\subsection{Twitter spam classification}

To check if there exist similar striking differences between spam and non-spam content posted by most active users on Twitter as well, we performed the same classification tasks on our Twitter data using a very similar feature set. Our Twitter dataset consisted of 23,555 spam tweets, and 198,974 non-spam tweets posted by 252 most active users during the 16 events. The feature set we used for tweets is shown in Table~\ref{tab:tw_feats}.

\begin{table}[!h]
\small
    \begin{tabular}{p{2.9cm}|p{4.9cm}}
    \hline
    Feature               & Description                                         \\ \hline
    tweet\_richness$^*$ & $\frac{tweet\_words}{tweet\_chars}$ \\
    num\_unique\_hashtags$^*$ & Number of unique hashtags in tweet                      \\
    num\_hashtags$^*$ & Total number of hashtags in tweet                      \\
    tweet\_chars & Number of characters in tweet                    \\
    num\_plain\_tokens & Number of words excluding hashtags, $@$mentions and URLs       \\
    tweet\_words & Total number of  words tweet                        \\
    tweet\_unique\_words & Number of unique words in tweet                  \\
    tweet\_source   &  Mobile / Web / 3$^{rd}$ party app / Other                  \\
    isRetweet   & True if tweet is a retweet; False otherwise      \\
    tweet\_rep\_factor & $\frac{tweet\_unique\_words}{tweet\_words}$ \\
    num\_urls$^*$   & Number of URLs in tweet                         \\
    hasMedia   &  True if tweet has picture / video; False otherwise         \\
    hashtag\_rep\_factor & $\frac{num\_unique\_hashtags}{num\_hashtags}$ \\
    num\_mentions   &  Number of $@$mentions in tweet       \\
    num\_unique\_mentions$^*$ & Number of unique $@$mentions in tweet                      \\
    mention\_rep\_factor & $\frac{num\_unique\_mentions}{num\_mentions}$ \\

 \hline
    \end{tabular}
\vspace{-8pt}
\caption{List of features used for spam classification on Twitter, in decreasing order of significance.}
\label{tab:tw_feats}
\vspace{-10pt}
\end{table}

Similar to Facebook, we were able to achieve a high accuracy of 98.2\% using the Random Forest classifier trained over this set of 16 features. This proved that there exist some obvious differences in the spam and non-spam content posted by most active users during real world events even in Twitter, unlike spam and non-spam content posted by common users in general. Table~\ref{tab:spam_classify} summarizes our classification results.

\begin{table}[!h]
\begin{center}
\begin{tabular}{p{2.5cm}|p{1.9cm}|p{1.4cm}|p{1.2cm}}
\hline
Classifier              & OSM (features) & Accuracy & F-Score \\ \hline
\multirow{3}{*}{Naive Bayesian}     & FB (21)         & 93.00\% & 0.927   \\ 
                        &   FB (15)       & 94.03\% & 0.945    \\ 
                        &   Twitter        &    90.67\%        &    0.874       \\ \hline
\multirow{3}{*}{J48 DecisionTree} & FB (21) & 99.82\%  & 0.998     \\ 
                        &   FB (15)       & 99.24\%  & 0.992    \\ 
                        &   Twitter        &    98.06\%        &    0.98       \\ \hline
\multirow{3}{*}{Random Forest}     & FB (21)         & {\bf 99.90}\%  & 0.999     \\ 
                        &   FB (15)       & 99.30\%  & 0.993    \\ 
                        &   Twitter        &    {\bf 98.20}\%        &    0.982       \\ \hline
\end{tabular}
\vspace{-7pt}
\caption{Classification results for Naive Bayesian, J48 Decision Tree, and Random Forest classifiers. Even with the 6 most important features removed, we were able to achieve an accuracy of over 99\% for Facebook.}
\label{tab:spam_classify}
\end{center}
\vspace{-20pt}
\end{table}

\begin{figure*}[!ht]
     \begin{center}
        \subfigure[Champions Trophy]{%
           \label{fig:ct_spam}
           \includegraphics[scale=0.3]{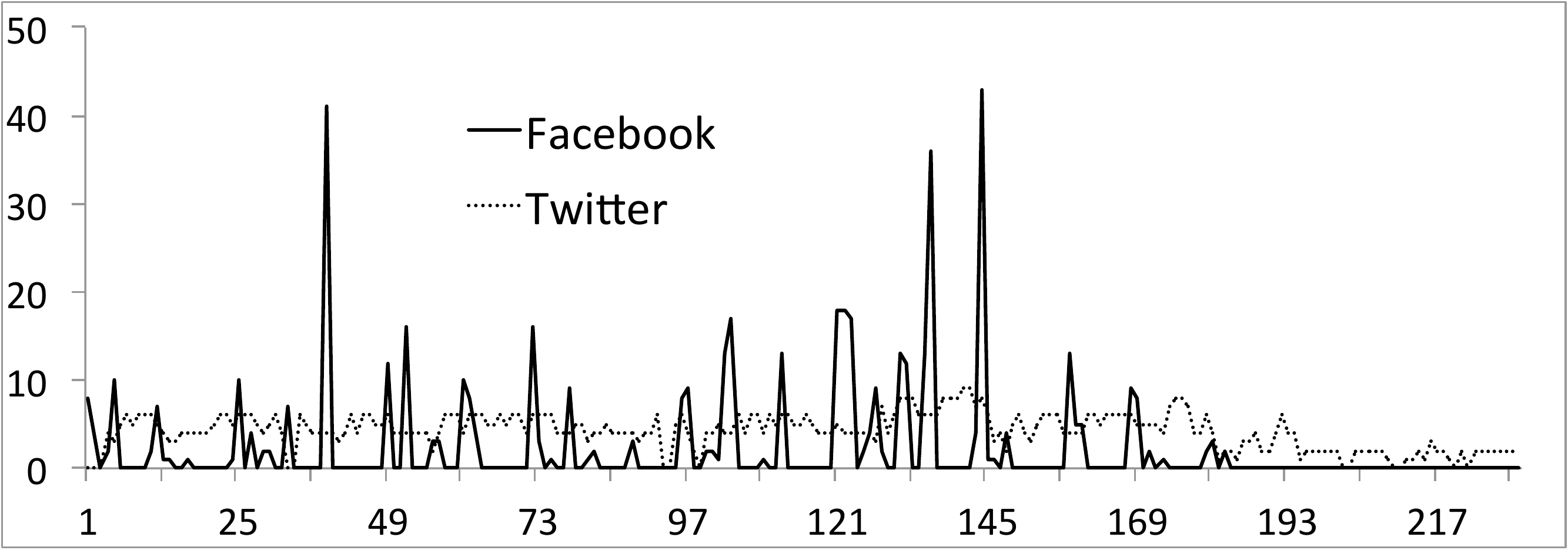}
        }
        	\subfigure[RIP Nelson Mandela]{%
            \label{fig:rip_spam}
            \includegraphics[scale=0.29]{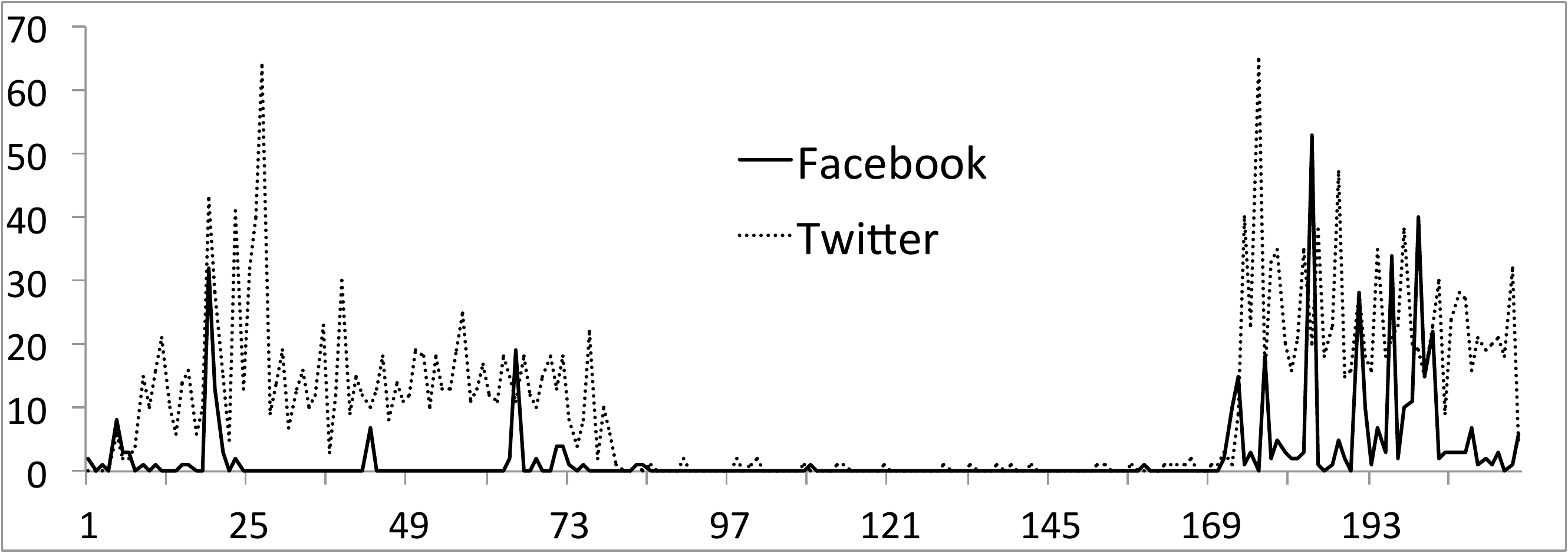}
        }
        	\subfigure[Royal Baby]{%
            \label{fig:royalbaby_spam}
            \includegraphics[scale=0.26]{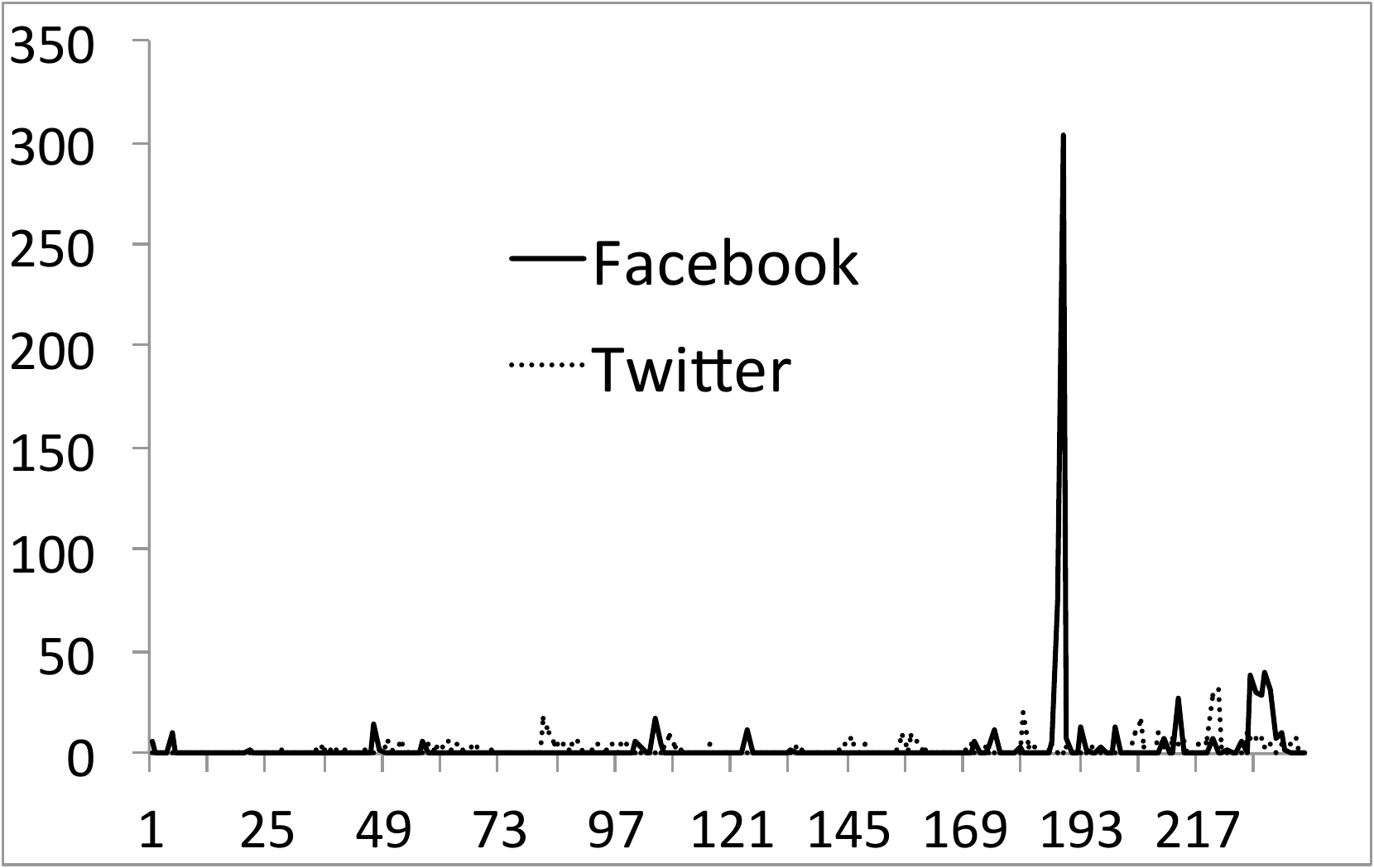}
        }
        	\subfigure[Telangana]{%
            \label{fig:telangana_spam}
            \includegraphics[scale=0.24]{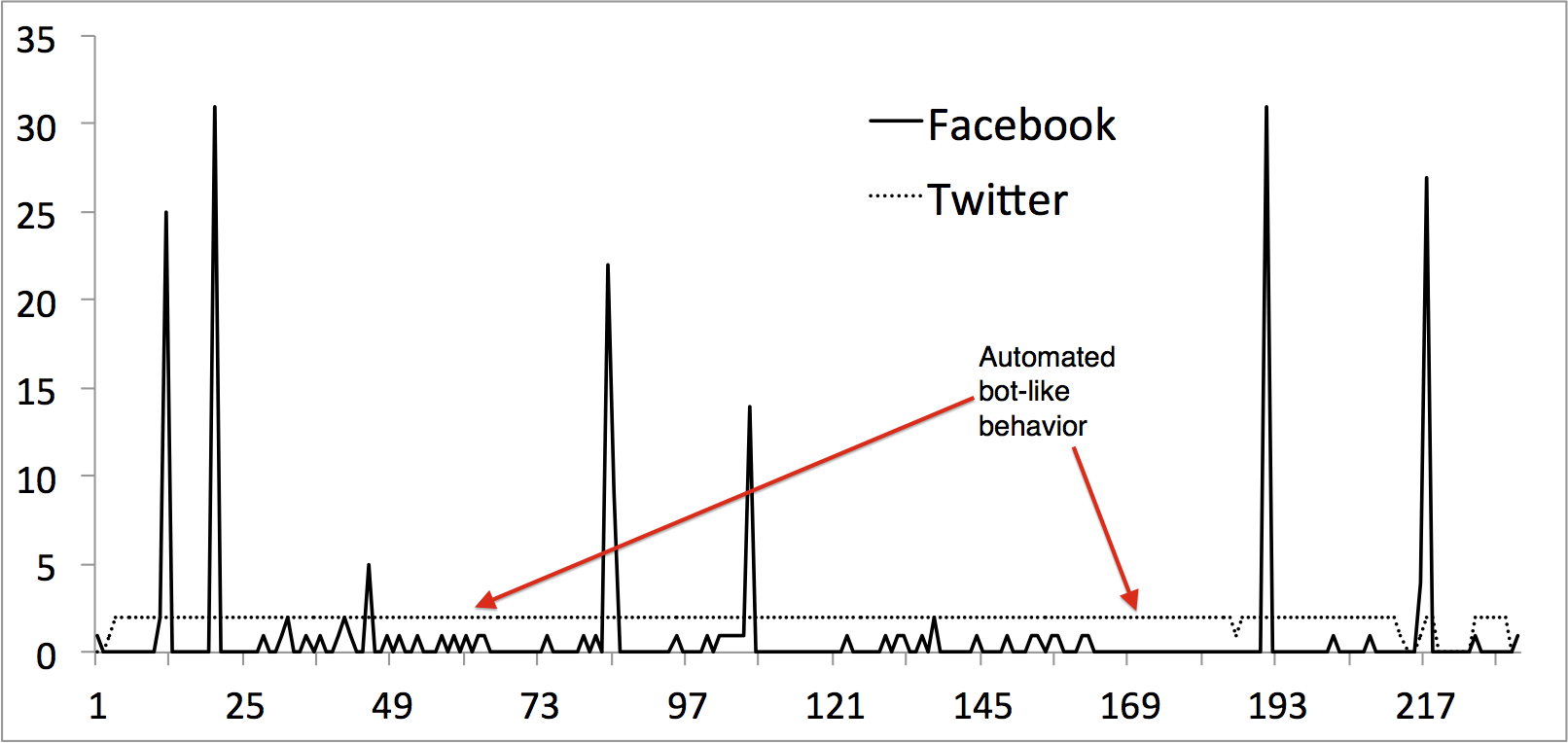}
        }
        	\subfigure[Typhoon Haiyan]{%
            \label{fig:th_spam}
            \includegraphics[scale=0.26]{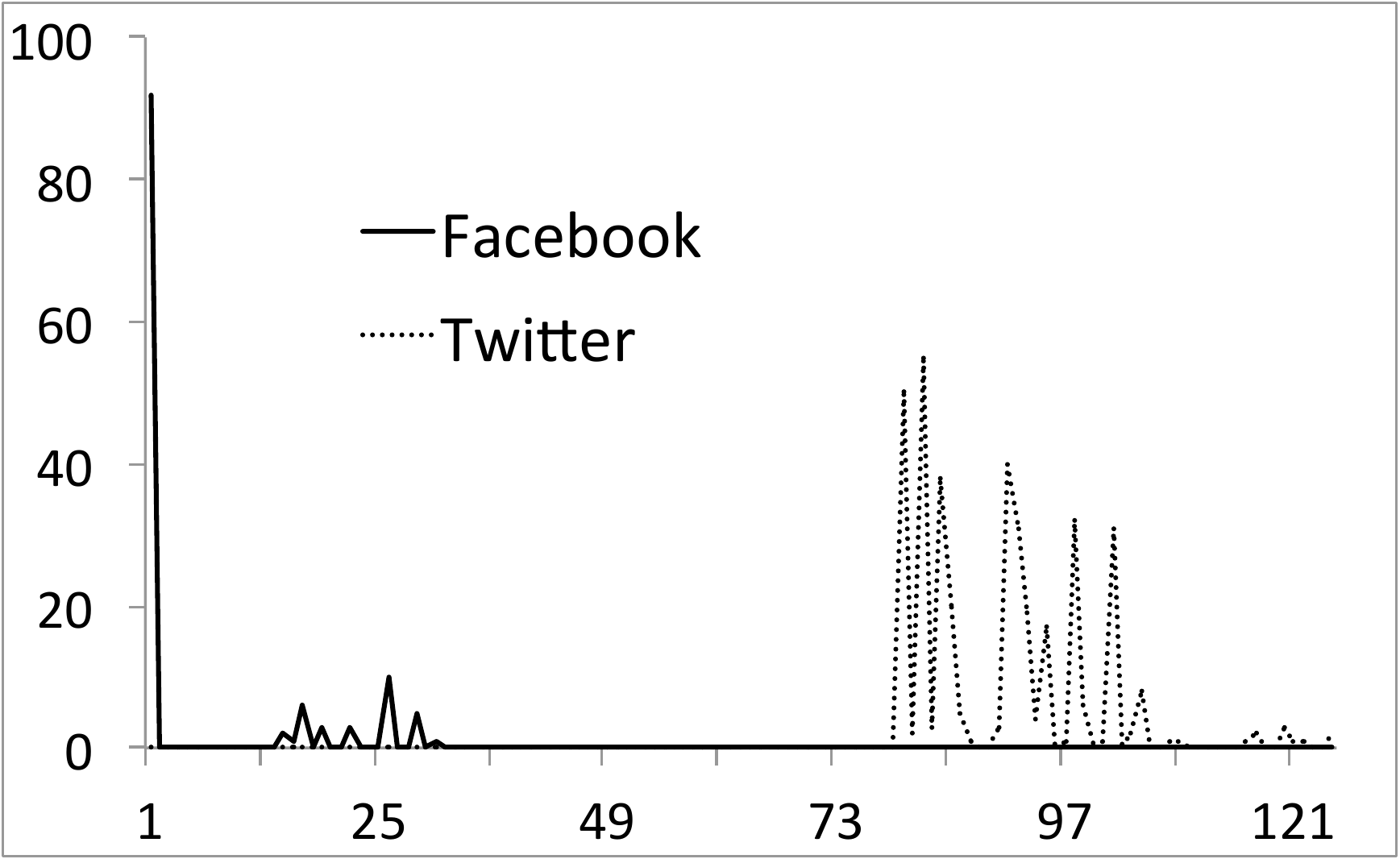}
        }
    \end{center}
\vspace{-17pt}
    \caption{%
Number of spam posts per hour, during 5 events for the first 10 days (240 hours) on Facebook and Twitter. Facebook depicted spiking behavior, while Twitter was less spiking, and fairly regular.
     }
   \label{fig:spam_time}
\vspace{-12pt}
\end{figure*}

\subsection{When do spammers post?}

Figure~\ref{fig:spam_time} shows the number of spam posts per hour on Facebook and Twitter, during the first 10 days (240 hours) of all the 5 events, where we found spam on Facebook. We found no correlation between the publish time of spam posts on Facebook and Twitter during any event (correl = 0.089; $\sigma$ = 0.013). Interestingly, during most events, we observed spiking behavior on Facebook. This possibly reflects that spammers on Facebook do not post spam regularly during an event; but post infrequently, and in bulk. On the other hand, spam on Twitter was observed to be fairly regular during most events. In fact, during the Telangana event (Figure~\ref{fig:telangana_spam}), we found highly automated behavior (most likely a bot account), where a user was posting a tweet after exactly 30 minutes (i.e. exactly 2 posts per hour consistently; as visible in the graph). Zhang et al. proposed techniques to study automated activity on Twitter~\cite{zhang2011detecting}, and found that accounts exhibiting such behavior were highly likely to be automated bots. 

The Nelson Mandela event witnessed two sub events; viz. Nelson Mandela's death on December 5 (hour 0), and his funeral on December 12 (hour 168). This is reflected by the dip in the graph in Figure~\ref{fig:rip_spam}, approximately between hour 80 and hour 168.

\section{Discussion and limitations} \label{sect:last}

In this paper, we presented a comparative analysis between Facebook and Twitter, and showed that there is a high overlap in the content of the two networks during real world events. We also showed that Facebook is quite fast at breaking and spreading content during real world events. Our analysis is based on a dataset containing publicly available content from Facebook and Twitter during 16 international events. The representativeness of this data is limited by the proportion of content which is public on these networks. 

It may be argued that the dynamics and purposes of Facebook totally differ from that of Twitter, and that the basis of such a comparison is questionable. However, we would like to point out that from a user's point of view, Facebook and Twitter are extremely similar. Both networks are timeline based, and users get to see content from only those entities which they choose (subscribed pages and friends on Facebook, and followed users on Twitter). On both networks, users can search for a topic, or click on a hashtag or trending topic (which is now also available on Facebook), to see what is being talked about the topic on the entire network as a whole. Users on both the networks, get to see only public content. Specially, when it comes to real-world events, one can argue that both Twitter and Facebook users are equally likely to learn (or not learn) about an event from the network, based on their interests and subscription.

One of the limitations of our work is the absence of Twitter data for the first few hours after an event occurs in the real world. This happened because the framework we used for collecting data, required manual input to start data collection for an event. We were thus able to initiate data collection for an event, only as soon as we learnt about the occurrence of the event. An alternate approach for data collection could have been to continuously collect a stream of tweets for trending topics, and then extract events from the stream. However, this approach is also bound to suffer from some delay, since it takes at least a couple of hours for an event to take place in the real world, and the topic to start trending on Twitter. To the best of our knowledge, there does not exist a way to be able to collect 100 per cent data for an event, unless the occurrence of the event is pre-determined, as in the case of events like IPL, Champions Trophy, birth of the Royal Baby etc.


To the best of our knowledge, this is the first attempt at comparing Facebook with Twitter content on such a large scale. No previous work has studied Facebook content during real world events. Apart from content in general, we analyzed the use of hashtags and URLs, and brought some interesting results to light, like presence of hashtags on Facebook before the launch of its official support by Facebook. We also highlighted how URLs posted on two networks at the same time can be used to connect two accounts on different social networks to the same real-world user.

In future, we intend to group together similar events in terms of geographic location, duration (short lived / long lasting), crisis / non-crisis, and expected / unexpected, and compare Facebook and Twitter behavior within, and across similar events. Due to space constraints, we could not add such analysis in this work. We would also like to explore ways of extracting network information from Facebook. This would help us in getting insights on community detection and content propagation during an event within the network, which have been previously studied on Twitter~\cite{gupta20131}. 

\bibliographystyle{abbrv}
\bibliography{Prateek}
\end{document}